\documentclass[aps,prl,reprint,groupedaddress]{revtex4-2} 
\usepackage[dvipsnames]{xcolor}
\usepackage[colorlinks=true,urlcolor=blue,citecolor=purple,linkcolor=red]{hyperref}
\usepackage{amssymb,amsmath,amsfonts}
\usepackage{epsfig}
\usepackage{graphicx}
\usepackage{epstopdf}
\usepackage{natbib}

\def\clock{{\count0=\time
		\divide\count0 60
		\ifnum\count0<10 0\fi\the\count0
		\multiply\count0 -60 \advance\count0 \time
		:\ifnum\count0<10 0\fi \the\count0
}}
\newcommand{\timestamp}{{\small\vbox{\hbox{\tt\jobname.tex}
			\hbox{\the\day/\the\month/\the\year, \clock}}}}

\newcommand{\nn}{\nonumber}

\newcommand{\be}{\begin{eqnarray}}
\newcommand{\ee}{\end{eqnarray}}
\newcommand{\beq}{\begin{eqnarray}}
\newcommand{\eeq}{\end{eqnarray}}

\newcommand{\beqa}{\begin{eqnarray}}
\newcommand{\eeqa}{\end{eqnarray}}
\newcommand{\D}{{\partial}}

\newcommand{\pd}[2]{\frac{\partial #1}{\partial #2}}

\newcommand{\diff}{{\text{d}}}

\let\oldsqrt\sqrt
\def\sqrt{\mathpalette\DHLhksqrt}
\def\DHLhksqrt#1#2{%
	\setbox0=\hbox{$#1\oldsqrt{#2\,}$}\dimen0=\ht0
	\advance\dimen0-0.2\ht0
	\setbox2=\hbox{\vrule height\ht0 depth -\dimen0}%
	{\box0\lower0.4pt\box2}}

\begin{document}

	\title{On the Non-Relativistic Expansion of Closed Bosonic Strings}

	\author{Jelle Hartong}
	\author{Emil Have}
	
	\email{j.hartong@ed.ac.uk}
	\email{Emil.Have@ed.ac.uk}

	\affiliation{School of Mathematics and Maxwell Institute for Mathematical Sciences,\\
		University of Edinburgh, Peter Guthrie Tait road, Edinburgh EH9 3FD, UK }

	

	\begin{abstract}
		We develop a novel approach to non-relativistic closed bosonic string theory that is based on a string $1/c^2$ expansion of the relativistic string, where $c$ is the speed of light. This approach has the benefit that one does not need to take a limit of a string in a near-critical Kalb--Ramond background. The $1/c^2$-expanded Polyakov action at next-to-leading order reproduces the known action of non-relativistic string theory provided that the target space obeys an appropriate foliation constraint. We compute the spectrum in a flat target space, with one circle direction that is wound by the string, up to next-to-leading order and show that it reproduces the spectrum of the Gomis--Ooguri string.
	\end{abstract}

	

	\maketitle



Non-relativistic string theory belongs to a growing class of string theories whose worldsheet and/or target spacetime is not described by a Lorentzian geometry. Such open and closed string theories allow us to study quantum gravity in non-Lorentzian domains, to embed non-Lorentzian field theories into a string context (e.g.~via world-volume theories of branes on which non-Lorentzian open strings end), to find new non-Lorentzian examples of holographic dualities, and to study interesting limits of standard string/M-theory. In this letter we focus on the particular case of non-relativistic strings.

The study of non-relativistic (NR) string theory began in earnest with the Gomis--Ooguri string \cite{Gomis:2000bd,Danielsson:2000gi}, which employs a near-critical Kalb--Ramond field to cancel a divergent term in the action and leads to a well-defined theory of strings in an infinite speed of light limit. In order for this theory to have a non-trivial spectrum, it was shown that the target space must have a circle direction that is wound by the string. Subsequent work centred on generalising the target space of NR string theory: in \cite{Harmark:2017rpg,Harmark:2018cdl}, NR strings were obtained via null reduction, followed by a duality transformation that replaced the null direction with a compact direction. In another direction, a theory of strings moving in a string Newton--Cartan background was developed in \cite{Bergshoeff:2018yvt}, which combined the limit approach of the original Gomis--Ooguri string with the notion of string Newton--Cartan geometry developed in \cite{Andringa:2012uz}. The relation between the null-reduced string and the string Newton--Cartan string was clarified in \cite{Harmark:2019upf}. Furthermore, NR strings have been shown to appear in double field theory, and doubled geometry in general turns out to include a wealth of non-Lorentzian geometries \cite{Ko:2015rha,Morand:2017fnv,Gallegos:2020egk}.

For point particles there are two ways to obtain a NR description starting from the relativistic one. Option one: we can start with the action of a massive particle and expand the geometry and the embedding scalars systematically in $1/c^2$ (see \cite{VandenBleeken:2017rij,Hansen:2020pqs}). Option two: we can place the particle in a near-critical electromagnetic field, choose the particle to be extremal by relating its charge and mass, and take a $c\rightarrow\infty$ limit \cite{Jensen:2014wha}. 
The latter approach is equivalent to performing a null reduction starting from a massless particle in one dimension higher. These two procedures in general do not lead to the same theory. In \cite{FSNC} option two is worked out for strings while in this letter we focus on option one.

As in the string Newton--Cartan geometry of \cite{Andringa:2012uz,Bergshoeff:2018yvt}, when performing a string $1/c^2$ expansion we single out not just the time direction, but also one spatial direction called the longitudinal target space direction. The target space becomes a string Newton--Cartan geometry that admits two-dimensional Lorentzian submanifolds.

The additional spatial direction singled out in the definition of the string Newton--Cartan geometry must be compact in order that the theory has a non-trivial spectrum \cite{Gomis:2000bd,Danielsson:2000gi,Danielsson:2000mu}. The circle provides a new length scale that can be compared with the string length. We will show that the non-relativistic expansion corresponds to radii that are much larger than the string length.

Our formalism allows us to formulate string theories at any given order of $1/c^2$. In this letter, we develop the formalism and demonstrate how the theory up to next-to-leading order (NLO) is related to existing NR strings, while the more elaborate next-to-next-leading (NNLO) theory will be considered in \cite{PNString}.

\noindent\textbf{Large-$c$ as a decompactification limit}.

It is a standard result of string theory that the mass squared of a quantum closed bosonic relativistic string in a 26-dimensional target space with a compact circle is
\be 
M^2 = \frac{\hbar^2 n^2}{c^2R^2} + \frac{w^2R^2}{{\alpha'}^2} + \frac{2}{\alpha'c}(N + \tilde{N} - 2\hbar)~,\label{eq:invariantmass}
\ee 
where $R$ is the radius of the circle, $n$ is the momentum mode and $w$ the winding number, while $N$ and $\tilde{N}$ are the number operators for the right and left movers. The dimensionless parameter with respect to which we will perform the NR expansion is $\epsilon = \alpha'\hbar/({cR^2})$. Taking $cT = T_{\text{eff}}$ and $R/c = R_{\text{eff}}$ to be independent of $c$, we obtain $\epsilon = \frac{\alpha'_{\text{eff}}\hbar}{c^2R^2_{\text{eff}}}$, where $\alpha'_{\text{eff}} = 1/(2\pi T_{\text{eff}})$. In this way, small values of $\epsilon$ correspond to large values of $R$, which leads us to conclude that the large $c$ limit in fact corresponds to a decompactification limit. More precisely, since the quantum of momentum in the compact direction is $\hbar/R$ and $R/\alpha'$ is the mass scale of the winding string, the center of mass velocity of the string along the compact direction is $v_{\text{com}} = \hbar\alpha'/R^2$, which is small compared to speed of light, $v_{\text{com}}/c\ll 1$, which we can equivalently interpret as a large $R$ limit. Since $E = \sqrt{M^2c^4 + p^2c^2}$, the mass in \eqref{eq:invariantmass} gives rise to
\be 
\hspace{-0.3cm}E &=& \frac{c^2  w R_{\text{eff}}}{\alpha'_{\text{eff}}} + \frac{N_{(0)} + \tilde N_{(0)}}{w R_{\text{eff}}} + \frac{\alpha'_{\text{eff}}}{2w R_{\text{eff}}}p_{(0)}^2 + \mathcal{O}(c^{-2})~,\label{eq:expansionofrelativisticenergy}
\ee 
where we have absorbed the normal ordering constant into $N_{(0)}$ and $\tilde N_{(0)}$ and where we have $1/c^2$-expanded the number operators and transverse momentum according to $X = X_{(0)} + \mathcal{O}(c^{-2})$
where $X = \{ N,\tilde{N}, p_i\}$. Equation \eqref{eq:expansionofrelativisticenergy} is the stringy version of the large-$c$ expansion of the point particle energy.

\noindent\textbf{String Newton--Cartan geometry}.

Paralleling the discussion of \cite{Hansen:2020pqs} (see also \cite{Bergshoeff:2018vfn,Bergshoeff:2019pij}), we now show how to obtain string Newton--Cartan geometry (SNC) from $D=d+2$-dimensional Lorentzian geometry. Write the Lorentzian metric $G_{MN}$ and its inverse as
\be 
\label{eq:metricdecomp}
\begin{aligned}
G_{MN} &= c^2\eta_{AB}T_M{^A}T_N{^B} +\Pi_{MN}^\perp~,\\
G^{MN} &= c^{-2} \eta^{AB}T^{M}{_A}T^N{_B}+\Pi^{\perp MN}~,
\end{aligned}
\ee 
where $M,N=0,1,\dots,d+1$ are spacetime indices, while $A,B=0,1$ are two-dimensional tangent space indices. We demand that $T{^M}_A\Pi_{MN}^\perp = T_M{^A}\Pi^{\perp MN} =0 $ and $ T{^M}_AT_M{^B}=\delta^B_A$. We expand these fields according to
\be 
\begin{aligned}
T_M{^A} &= \tau_M{^A}+c^{-2}m_M{^A}+\mathcal{O}(c^{-4})~,\\
\Pi_{MN}^\perp &= H_{MN}^\perp + \mathcal{O}(c^{-2})~,
\end{aligned}
\ee 
with similar expansions for $T^M{_A}$ and $\Pi^{\perp MN}$. We then find that the metric expands as
\be 
G_{MN}=c^2\tau_{MN} + H_{MN} 
+\mathcal{O}(c^{-2})~,\label{eq:expofmetric}
\ee 
where $\tau_{MN}=\eta_{AB}\tau_M{^A}\tau_N{^B}$ and $H_{MN} = H^\perp_{MN} + 2\eta_{AB} \tau_{(M}{^A}m_{N)}{^B}$.

We define the \textit{strong foliation constraint} in terms of the 1-forms $\tau^A=\tau_M{}^Adx^M$ as
\be\label{eq:strong-foliation-constraint}
d\tau^A=\omega\varepsilon^A{}_B\wedge\tau^B\,,
\ee
where $\omega$ is a 1-form that is determined by solving 
\eqref{eq:strong-foliation-constraint} for $\omega$. However, equation \eqref{eq:strong-foliation-constraint} does more than determining $\omega$ since it also constrains $\tau^A$. The constraint \eqref{eq:strong-foliation-constraint} played an important role in \cite{Bergshoeff:2018yvt} in their definition of NR string theory. This condition has recently been relaxed in \cite{Yan:2021lbe,Bergshoeff:2021bmc}. Here we will show that the $1/c^2$ expansion naturally comes with its own foliation constraint
\footnote{In \cite{PNString} we will show that the string $1/c^2$ expansion of a relativistic geometry naturally leads to what we call a type II torsional SNC or TSNC geometry in analogy with the particle $1/c^2$ expansion studied in \cite{Hansen:2018ofj,Hansen:2020pqs}. Upon using the strong foliation constraint (\ref{eq:strong-foliation-constraint}) this reduces to an ordinary SNC geometry as developed in e.g.~\cite{Bergshoeff:2019pij}}.

\noindent\textbf{Codimension-two foliations.}

The target space fields of the relativistic string obey the beta function equations involving the metric, the Kalb--Ramond 2-form and the dilaton. If we ignore the dilaton and the 2-form, these equations are simply the Einstein equations $R_{MN}=0$ to leading order in $\alpha'$. If we expand these in $1/c^2$ using \eqref{eq:metricdecomp}, we find at leading order (LO) that
\begin{equation}
H^{\perp QS}H^{\perp RT}\left(d\tau^A\right)_{QR}\left(d\tau^B\right)_{ST}=0~,
\end{equation}
where $H^{\perp MN}$ is the leading-order component of $\Pi^{\perp MN}$. The above is a sum of squares for $A=B=0,1$ and thus equivalent to $H^{\perp QS}H^{\perp RT}\left(d\tau^A\right)_{QR}=0$. This in turn is equivalent to
\begin{equation}
\label{eq:Frobenius}
    d\tau^A=\alpha^A{}_B\wedge \tau^B\,,
\end{equation}
for arbitrary one-forms $\alpha^A{}_B$. This is the Frobenius integrability condition for a codimension-two foliation whose leaves are $d$-dimensional Riemannian spaces with normal one-forms $\tau^A$. This reduces to the strong foliation constraint \eqref{eq:strong-foliation-constraint} only if $\alpha{^A}{_B} = \omega \varepsilon^A{_B}$. Equation \eqref{eq:Frobenius} is the string NC analogue of the TTNC condition imposed in NC geometry \cite{Christensen:2013lma} which likewise follows from the particle $1/c^2$ expansion of the Einstein equations \cite{Hansen:2018ofj,VandenBleeken:2017rij}.

\noindent\textbf{Expansion of the string action}.

The Polyakov Lagrangian is
\be 
\label{eq:RelativisticPolyakov}
\mathcal{L}_{\text{P}}=- \frac{cT}{2} \sqrt{-\gamma}\gamma^{\alpha\beta}\D_\alpha X^M \D_\beta X^N G_{MN}~.
\ee
To expand this, we must, in addition to expanding the metric $G_{MN}$ as in \eqref{eq:expofmetric}, expand the embedding field (that will in general depend on $c$) as $X^M = x^M + c^{-2}y^M + \mathcal{O}(c^{-4})$. We also expand the worldsheet metric $\gamma_{\alpha\beta}$ as
\be \label{eq:expgamma}
\gamma_{\alpha\beta} =  \gamma_{(0)\alpha\beta} +c^{-2} \gamma_{(2)\alpha\beta} + \mathcal{O}(c^{-4})~,
\ee 
where $\gamma_{(0)\alpha\beta}$ is a Lorentzian metric with determinant $\sqrt{-\det\gamma_{(0)}} = e$,
while $\gamma_{(2)\alpha\beta}$ is a symmetric tensor. The pullback $G_{\alpha\beta}(X) = \D_\alpha X^M\D_\beta X^NG_{MN}(X)$ acquires the following expansion
\be 
G_{\alpha\beta}(X) = c^2\tau_{\alpha\beta}(x) + H_{\alpha\beta}(x,y) + \mathcal{O}(c^{-2})~,
\ee 
where $\tau_{\alpha\beta}(x) = \D_\alpha x^M\D_\beta x^N\tau_{MN}(x)$ is assumed to be of Lorentzian signature and where 
\begin{eqnarray}
H_{\alpha\beta}(x,y) &=& \partial_\alpha x^M\partial_\beta x^N H_{MN}(x) + 2\D_{(\alpha} x^M \D_{\beta)}y^N\tau_{MN}(x)\nn\\
&&+ \D_\alpha x^M\D_\beta x^N y^L \D_L \tau_{MN}(x)~.
\end{eqnarray}

The $1/c^2$ expansion of the Polyakov Lagrangian \eqref{eq:RelativisticPolyakov} is $\mathcal{L}_{\text{P}} = c^2\mathcal{L}_{\text{P-LO}} + \mathcal{L}_{\text{P-NLO}} + \mathcal{O}(c^{-2})$, where
\begin{align}
\mathcal{L}_{\text{P-LO}} =&  -\frac{T_{\text{eff}}}{2} e \gamma_{(0)}^{\alpha\beta}\tau_{\alpha\beta}~,\label{eq:LOPLagrangian}\\
\begin{split}
\mathcal{L}_{\text{P-NLO}} =& -\frac{T_{\text{eff}}}{2}e\gamma_{(0)}^{\alpha\beta} H_{\alpha\beta} + \frac{T_{\text{eff}}}{4}e G_{(0)}^{\alpha\beta\gamma\delta} \tau_{\alpha\beta}\gamma_{(2)\gamma\delta}\\
&+ y^M\frac{\delta \mathcal{L}_{\text{P-LO}}}{\delta x^M}~,\label{eq:NLOPLagrangian}
\end{split}
\end{align} 
where we introduced the Wheeler--DeWitt metric $G_{(0)}^{\alpha\beta\gamma\delta} = \gamma_{(0)}^{\alpha\gamma}\gamma_{(0)}^{\delta\beta} + \gamma_{(0)}^{\alpha\delta}\gamma_{(0)}^{\gamma\beta}- \gamma_{(0)}^{\alpha\beta}\gamma_{(0)}^{\gamma\delta}$.
The reason for expanding the worldsheet metric as in \eqref{eq:expgamma} is so that $\gamma_{(0)\alpha\beta}$ can be related to the Lorentzian pullback metric $\tau_{\alpha\beta}$ via the equation of motion of $\gamma_{(0)\alpha\beta}$.

The NLO theory can be recast in a Nambu--Goto (NG) formulation. By integrating out $\gamma_{(0)\alpha\beta}$ from the P-LO Lagrangian, and by integrating out both $\gamma_{(0)\alpha\beta}$ and $\gamma_{(2)\alpha\beta}$ from the P-NLO Lagrangian, the NG Lagrangian at LO and NLO can be found to be
\be 
\label{eq:NG-Lagrangians}
\begin{aligned}
    \mathcal{L}_{\text{NG-LO}} &= -T_{\text{eff}} \sqrt{-\tau}~,\\
 \mathcal{L}_{\text{NG-NLO}} &= -\frac{T_{\text{eff}}}{2} \sqrt{-\tau}\tau^{\alpha\beta}H_{\alpha\beta}(x) + y^M\frac{\delta \mathcal{L}_{\text{NG-LO}}}{\delta x^M} ~,
\end{aligned}
\ee 
where $\tau = \det \tau_{\alpha\beta}$. The constraints that come from integrating out $\gamma_{(0)\alpha\beta}$ and $\gamma_{(2)\alpha\beta}$ are the LO and NLO Virasoro constraints, respectively. These can also be obtained by $1/c^2$-expanding the Virasoro constraints obtained by integrating out $\gamma_{\alpha\beta}$ in \eqref{eq:RelativisticPolyakov}. We remark that the constraint from integrating out $\gamma_{(2)\alpha\beta}$ at NLO is identical to that from integrating out $\gamma_{(0)\alpha\beta}$ at LO and leads to
\begin{equation}\label{eq:ConstraintLO}
    \tau_{\alpha\beta}=\frac{1}{2}\gamma_{(0)}^{\gamma\delta}\tau_{\gamma\delta}\gamma_{(0)\alpha\beta}\,.
\end{equation}

The equation of motion of the NG-LO Lagrangian for the embedding scalar $x^M$, which features in \eqref{eq:NG-Lagrangians}, reads
\begin{eqnarray}
 \frac{\delta \mathcal{L}_{\text{NG-LO}}}{\delta x^M} &=& -T_{\text{eff}}
   \varepsilon^{\alpha\beta}\varepsilon_{AB}\D_\alpha x^K\D_\beta x^L\left[\tau_M{^A}\D_{[K}\tau_{L]}{^B}\right.\nn\\
   &&\left.-2\tau_L{^B}\D_{[M}\tau_{K]}{^A}\right]\,.\label{eq:LO-eom}
\end{eqnarray}
If we assume that the target space obeys the Frobenius condition \eqref{eq:Frobenius}, then equation \eqref{eq:LO-eom} forces $\alpha_M{}^A{}_A$ to be equal to $\tau_M{}^A X_A$ for some $X_A$ that is determined by \eqref{eq:LO-eom}. A sufficient condition for \eqref{eq:LO-eom} to be equal to zero (and hence for the $y$-term to drop out of $\mathcal{L}_{\text{NG-NLO}}$) is to simply take $\alpha_M{}^A{}_B$ to be traceless of which the strong foliation constraint \eqref{eq:strong-foliation-constraint} is a special case.

\noindent\textbf{Relation with the NR string action}.

We now make contact with the NR string actions presented in \cite{Harmark:2017rpg,Bergshoeff:2018yvt,Harmark:2019upf}. Writing $\gamma_{(0)\alpha\beta} = \eta_{ab}e_\alpha{^a}e_\beta{^b}$ where $a,b=0,1$ are tangent space worldsheet indices, and $\gamma_{(2)\alpha\beta} =  2e_{(\alpha}{^a}e_{\beta)}{^b}A_{ab}$ for some symmetric $A_{ab}$, the Lagrange multiplier term involving $\gamma_{(2)}$ in \eqref{eq:NLOPLagrangian} is
\be 
\mathcal{L}_{\text{LM}} &=& \frac{T_{\text{eff}}}{2}e G_{(0)}^{abcd}e^\alpha{_a}\left(-\tau_\alpha{^0}+\tau_\alpha{^1} \right)e^\beta{_b}\left(\tau_\beta{^0} + \tau_\beta{^1}  \right) A_{cd}~,\nn
\ee 
where $G_{(0)}^{abcd} =\eta^{ac}\eta^{bd} + \eta^{ad}\eta^{bc}- \eta^{ab}\eta^{cd}$. In worldsheet tangent space lightcone coordinates with $\eta_{+-}=-1/2$ and $\eta_{++}=\eta_{--}=0$, the WDW metric has only two non-zero components $G_{(0)}^{++--} = G_{(0)}^{--++} = 8$, implying that $\mathcal{L}_{\text{LM}}$ imposes two constraints via $A_{++}$ and $A_{--}$,
\be 
\begin{aligned}
    0&= e^\alpha{_-}\left(-\tau_\alpha{^0}+\tau_\alpha{^1} \right)e^\beta{_-}\left(\tau_\beta{^0} + \tau_\beta{^1}  \right)~,\\
 0&=e^\alpha{_+}\left(-\tau_\alpha{^0}+\tau_\alpha{^1} \right)e^\beta{_+}\left(\tau_\beta{^0} + \tau_\beta{}^1  \right)~.
\end{aligned}
\ee 
The $e^\alpha{}_\pm$ projections of  $\left(-\tau_\alpha{}^0+\tau_\alpha{^1} \right)$ cannot both be zero and likewise for $\left(\tau_\alpha{^0}+\tau_\alpha{^1} \right)$ because by completeness of the $e^\alpha{}_\pm$ this would imply that $\tau_\alpha{^A}$ is not invertible. Without loss of generality we can assume that
\be 
\hspace{-0.3cm}e^\alpha{_-}\left(-\tau_\alpha{^0}+\tau_\alpha{^1} \right) \neq 0\,,\qquad e^\beta{_+}\left(\tau_\beta{^0} + \tau_\beta{^1}  \right)\neq 0~.
\ee 
Hence, we can make the redefinitions  $\tilde\lambda_\pm = 4e^\alpha{_\mp}\left(\mp\tau_\alpha{^0}+\tau_\alpha{^1} \right)A_{\pm\pm}$, leading to
\be \label{eq:LM}
\hspace{-0.3cm}\mathcal{L}_{\text{LM}}  &=& -\frac{T_{\text{eff}}}{2}\left[ \tilde\lambda_+\varepsilon^{\alpha\beta}e_\alpha{^+}\tau_\beta{^+} +\tilde\lambda_-\varepsilon^{\alpha\beta}e_\alpha{^-}\tau_\beta{^-} \right]~,
\ee 
where we defined $\tau_\beta{^\pm} = \tau_\beta{^0} \pm \tau_\beta{^1}$. 

The final term in \eqref{eq:NLOPLagrangian} can be related to the variation of the NG action at LO via the following identity
\begin{eqnarray}
\label{eq:identity-for-eoms}
\begin{aligned}
&\hspace{-0.3cm}y^M\frac{\delta \mathcal{L}_{\text{P-LO}}}{\delta x^M}  =  y^M\frac{\delta \mathcal{L}_{\text{NG-LO}}}{\delta x^M}\\
&\hspace{-0.3cm}-T_{\text{eff}}\varepsilon^{\alpha\gamma}\tau_\alpha{^+}e_\gamma{^+}e^\beta{_+}\left(\tau_M{^-}\partial_\beta y^M+y^M\partial_\beta x^N \partial_M\tau_N{^-}\right)\\
&\hspace{-0.3cm}+T_{\text{eff}}\varepsilon^{\alpha\gamma}\tau_\alpha{^-}e_\gamma{^-}e^\beta{_-}\left(\tau_M{^+}\partial_\beta y^M+y^M\partial_\beta x^N \partial_M\tau_N{^+}\right)~.
\end{aligned}
\end{eqnarray}
Using this identity and \eqref{eq:LM} we can rewrite \eqref{eq:NLOPLagrangian} as
\be 
\label{eq:NLOPLagrangian-SNC}
\begin{aligned}
\mathcal{L}_{\text{P-NLO}} &=
-\frac{T_{\text{eff}}}{2}\varepsilon^{\alpha\beta}\left[ \lambda_+e_\alpha{^+}\tau_\beta{^+} +\lambda_-e_\alpha{^-}\tau_\beta{^-} \right]\\
&-\frac{T_{\text{eff}}}{2}e \gamma^{\alpha\beta}_{(0)}H_{\alpha\beta}+y^M\frac{\delta \mathcal{L}_{\text{NG-LO}}}{\delta x^M}~,
\end{aligned}
\ee 
where we defined $\lambda_\pm = \tilde\lambda_\pm  \mp 2 e^\gamma{_\pm}(\tau_N{^\mp}\D_\gamma y^N + \D_\gamma x^M y^N \partial_N \tau_M{^\mp} )$. This reduces to the SNC string Lagrangian of \cite{Harmark:2017rpg,Bergshoeff:2018yvt,Harmark:2019upf} iff equation \eqref{eq:LO-eom} vanishes. A sufficient condition for this to be the case is to set $\alpha^A{}_A = 0$.

\noindent\textbf{The Wess--Zumino term}.

We can add a Kalb--Ramond field via the Wess--Zumino (WZ) Lagrangian
\be 
\label{eq:WZ-term}
\mathcal{L}_{\text{WZ}} &=& -\frac{cT}{2}\varepsilon^{\alpha\beta} \D_\alpha X^M \D_\beta X^N B_{MN}(X)~.
\ee 
Expanding $B_{MN} = c^2B_{(-2)MN} + B_{(0)MN} + \mathcal{O}(c^{-2})$, we find that $\mathcal{L}_{\text{WZ}} = c^2 \mathcal{L}_{\text{WZ-LO}} + \mathcal{L}_{\text{WZ-NLO}} + \mathcal{O}(c^{-2})$, where 
\be 
\begin{aligned}
\mathcal{L}_{\text{WZ-LO}} &= -\frac{T_{\text{eff}}}{2}\epsilon^{\alpha\beta} B_{(-2)\alpha\beta}~,\\
\mathcal{L}_{\text{WZ-NLO}} &= -\frac{T_{\text{eff}}}{2}\epsilon^{\alpha\beta} B_{(0)\alpha\beta} + y^M\frac{\delta \mathcal{L}_{\text{WZ-LO}}}{\delta x^M}~.
\end{aligned}
\ee 
In the presence of a Kalb--Ramond field, the LO equation of motion \eqref{eq:LO-eom} picks up the additional term
\be 
\frac{\delta \mathcal{L}_{\text{WZ-LO}}}{\delta x^M} &=&  -\frac{T_{\text{eff}}}{2}\epsilon^{\alpha\beta}\D_\alpha x^N \D_\beta x^L H_{(-2)MNL}~,
\ee 
where $H_{(-2)MNL} = 3\D_{[M}B_{(-2)NL]}$ is the field strength of $B_{(-2)MN}$. As was observed in \cite{Gomis:2000bd,Danielsson:2000gi,Bergshoeff:2018yvt}, it is possible to choose the LO $B$ field such that the LO Lagrangians  
$\mathcal{L}_{\text{WZ-LO}}$ and $\mathcal{L}_{\text{NG-LO}}$ cancel.

\noindent\textbf{Spectrum in flat space}.

For the remainder of this letter we consider a flat string NC geometry for which $\tau_M{^0}= \delta_M^t$, $\tau_M{^1} = \delta^v_M$, $m_M{^A} = 0$, $H_{MN} = \delta^i_M\delta^i_N$ with $i = 2,\dots,d+1$ a transverse index.

In the relativistic Polyakov action the worldsheet reparametrisations are generated by, say, $\Xi^\alpha$ which acts infinitesimally on $\gamma_{\alpha\beta}$ and $X^M$ via Lie derivatives along $\Xi^\alpha$. Expanding both the fields and the parameters $\Xi^\alpha=\xi_{(0)}^\alpha+c^{-2}\xi_{(2)}^\alpha+\mathcal{O}(c^{-4})$ we learn that $x^M$ and $y^M$ transform as $\delta x^M=\xi^\alpha_{(0)}\partial_\alpha x^M$ and $\delta y^M=\xi^\alpha_{(0)}\partial_\alpha y^M+\xi^\alpha_{(2)}\partial_\alpha x^M$ and likewise for $\gamma_{(0)\alpha\beta}$ and $\gamma_{(2)\alpha\beta}$. The latter also transform under the $1/c^2$-expanded Weyl transformations $\delta\gamma_{\alpha\beta}=2\omega\gamma_{\alpha\beta}$ where $\omega=\omega_{(0)}+c^{-2}\omega_{(2)}+\mathcal{O}(c^{-4})$. We can gauge fix the LO and NLO worldsheet gauge redundancy generated by $\xi_{(0)}^\alpha$, $\xi_{(2)}^\alpha$, $\omega_{(0)}$, $\omega_{(2)}$, locally by setting $\gamma_{(0)\alpha\beta} = \eta_{\alpha\beta}$ and $\gamma_{(2)\alpha\beta} =0$. The residual gauge transformations acting on $x^M$, $y^M$ are generated by  $\xi_{(n)}=\xi^+_{(n)}(\sigma^+)\partial_++\xi^-_{(n)}(\sigma^-)\partial_-$ for $n=0,2$ where all functions of $\sigma^\pm=\sigma^0\pm\sigma^1$ are periodic.

The gauge-fixed P-LO Lagrangian on flat space is
\be 
\mathcal{L}_{\text{P-LO}} &=& \frac{T_{\text{eff}}}{2}\eta^{\alpha\beta}\D_\alpha x^t\D_\beta x^t - \frac{T_{\text{eff}}}{2}\eta^{\alpha\beta} \D_\alpha x^v\D_\beta x^v~.
\ee 
The Virasoro constraints from integrating out $\gamma_{(0)\alpha\beta}$ in $\mathcal{L}_{\text{P-LO}}$ are \eqref{eq:ConstraintLO}, which amount to $\tau_{++}=0=\tau_{--}$. Without loss of generality this is equivalent to the LO constraints: $\D_-x^+ = 0$ and $\D_+x^-=0$, where we defined $x^\pm = x^t \pm x^v$. In our conventions $x^v$ has dimensions of time. 
Since the $v$-direction is compact, the constraints $\D_\mp x^\pm = 0$ imply the following mode expansions for $x^\pm$
\be 
x^\pm &=& x_0^\pm + w R_{\text{eff}} \sigma^\pm + \sigma^\pm\text{-oscillations}~,
\ee 
where $x_0^\pm$ are constants, $w$ is the winding number and $R_{\text{eff}}$ is the target space circle radius (in units of time). The residual LO diffeomorphisms $\xi^\pm_{(0)}(\sigma^\pm)$ act on $x^\pm$ as $\delta_{\xi_{(0)}}x^\pm = \xi^\pm_{(0)}\D_\pm x^\pm$, where we have used the LO constraints  
and since $\xi_{(0)}^\pm$ is periodic, we can fix the residual gauge transformations by removing the non-zero modes of $x^\pm$, leaving only $x^\pm = x_0^\pm + w R_{\text{eff}}\sigma^\pm$.

The relativistic energy expands in $1/c^2$ as
\be 
E &=& c^2 E_{\text{LO}} + E_{\text{NLO}} + \mathcal{O}(c^{-2})\\
&=& -c^2\oint\diff\sigma^1\pd{\mathcal{L}_{\text{P-LO}}}{\D_0x^t} - \oint\diff\sigma^1\pd{\mathcal{L}_{\text{P-NLO}}}{\D_0x^t}  +\mathcal{O}(c^{-2})~.\nn
\ee 
Hence, the energy at LO is the `stringy' rest mass 
\be 
\hspace{-0.3cm}E_{\text{LO}} &=& -\oint\diff\sigma^1\frac{\D \mathcal{L}_{\text{P-LO}}}{\D(\D_0x^t)} = \frac{wR_{\text{eff}}}{\alpha'_{\text{eff}}}~,\label{eq:LOenergy}
\ee 
which matches the first term in \eqref{eq:expansionofrelativisticenergy}.

If we include a Kalb--Ramond field of the form $B_{MN} = 2c^2\delta^t_{[M}\delta^v_{N]}(1-\lambda)$, corresponding to $B_{(-2)MN} = 2\delta^t_{[M}\delta^v_{N]}(1-\lambda)$ and $B_{(0)MN} =0$, we produce the `instanton term' of \cite{Gomis:2000bd} for $\lambda = 1/2$ (see also \cite{Danielsson:2000mu})
\be 
E_{\text{LO}} &=& \frac{\lambda wR_{\text{eff}}}{\alpha'_{\text{eff}}}  ~.\label{eq:NLOspectrum+WZ}
\ee 
In what follows, we will take $\lambda = 1$.

The gauge-fixed P-NLO Lagrangian on flat space is
\be
\begin{aligned}
 \mathcal{L}_{\text{P-NLO}} &= -\frac{T_{\text{eff}}}{2}\eta^{\alpha\beta}\D_\alpha x^i\D_\beta x^i\\
 &+T_{\text{eff}}\eta^{\alpha\beta}\D_\alpha y^t\D_\beta x^t - T_{\text{eff}}\eta^{\alpha\beta}\D_\alpha y^v\D_\beta x^v~,
 \end{aligned}
 \ee 
where the equations of motion for $y^t$ and $y^v$, respectively, imply the LO equations of motion for $x^t$ and $x^v$, while the NLO equations of $x^t$ and $x^v$ tell us that $\D_+\D_- y^t = \D_+\D_- y^v =0 $. The equation of motion for $x^i$ implies that $\D_+\D_-x^i = 0$. This leads to the mode expansions: \footnote{We are excluding a term linear in $\sigma^1$ in $y^v$. If we had not done so, this would have corresponded to expanding the winding number as $w = w_{(0)} + c^{-2}w_{(2)} + \mathcal{O}(c^{-4})$, where $w_{(2)}$ is a subleading winding number appearing in the mode expansion of $y^v$. Since $w$ is an integer we choose not to include a term linear in $\sigma^1$ in $y^v$.}
\begin{align}
\begin{split}
x^i &= x_0^i + \frac{1}{2\pi T_{\text{eff}}}p_{(0)i} \sigma^0\\
&+ \frac{1}{\sqrt{4\pi T_{\text{eff}}}}\sum_{k\neq 0} \frac{i}{k}\left[\alpha_k^i e^{-ik\sigma^-}+\tilde{\alpha}^i_k e^{-ik\sigma^+} \right]~,
\end{split}\\
y^\pm &= y^\pm_0 - \frac{1}{2\pi T_{\text{eff}}}p_{(0)\mp}(\sigma^++\sigma^-) + \text{oscillations}~,\label{eq:ypm}
\end{align}
where $y^\pm = y^t \pm y^v$, and where the momenta $p_{(0)\pm}$ are the canonical momenta, $p_{(0)\pm} = \oint\diff\sigma^1\frac{\D\mathcal{L}_{\text{P-NLO}}}{\D(\D_0 x^\pm)}$. The constraints arising from integrating out $\gamma_{(2)\alpha\beta}$ from \eqref{eq:NLOPLagrangian} are the same as those originating from integrating out $\gamma_{(0)\alpha\beta}$ from \eqref{eq:LOPLagrangian}. The constraints from integrating out $\gamma_{(0)\alpha\beta}$ from \eqref{eq:NLOPLagrangian} read
\be 
\D_\mp y^\pm = \frac{1}{w R_{\text{eff}}}\D_\mp x^i\D_\mp x^i~.\label{eq:NLOconstraints}
\ee

We still need to fix the subleading residual gauge invariance $\xi_{(2)}^\pm(\sigma^\pm)$, which acts infinitesimally on $y^\pm$ as $\delta y^\pm =  w R_{\text{eff}}\xi_{(2)}^\pm(\sigma^\pm)$. We fix the subleading residual gauge transformations by removing the oscillations in $\D_\pm y^\pm$.

The zero mode of each Virasoro constraint in \eqref{eq:NLOconstraints} gives an expression for $p_{(0)\pm} = \oint\diff\sigma^1\pd{\mathcal{L}_{\text{NLO-P}}}{(\D_0 x^\pm)} = \tfrac12(p_{(0)t} \pm p_{(0)v})$, respectively, in terms of the modes of $x^i$, 
\begin{align} 
\label{eq:expressions-for-p-from-Virasoro}
\begin{split}
p_{(0)-} &= -\frac{N_{(0)}}{wR_{\text{eff}}} - \frac{\alpha'_{\text{eff}}}{4wR_{\text{eff}}}(p_{(0)})^2~,\\ p_{(0)+} &= -\frac{\tilde N_{(0)}}{wR_{\text{eff}}} - \frac{\alpha'_{\text{eff}}}{4wR_{\text{eff}}}(p_{(0)})^2~,
\end{split}
\end{align}
where we defined the leading number operators $N_{(0)} = \sum_{n=1}^\infty \alpha_{-n}^i \alpha_n^i$ and $\tilde N_{(0)} = \sum_{n=1}^\infty \tilde\alpha_{-n}^i \tilde\alpha_n^i$. Adding the expressions in \eqref{eq:expressions-for-p-from-Virasoro}, we get
\be
\hspace{-1cm}E_{\text{NLO}} &=& -p_{(0)t}
=  \frac{N_{(0)} + \tilde N_{(0)}}{wR_{\text{eff}}} + \frac{\alpha'_{\text{eff}}}{2w R_{\text{eff}}}p_{(0)}^2~,\label{eq:NLOspectrum}
\ee 
in agreement with the $c^0$ terms in \eqref{eq:expansionofrelativisticenergy}. 

The momentum in the compact $v$-direction is quantised, $p_{(0)v} = \tfrac{\hbar n}{R_{\text{eff}}}$,  where $n$ is an integer. Since $p_{(0)v}=p_{(0)+}-p_{(0)-}$, we obtain the level matching condition
$N_{(0)} - \tilde N_{(0)} = \hbar nw$. Canonically quantising the NLO theory will lead to a normal ordering constant $a\hbar$ in the number operators  $N_{(0)} = \sum_{n=1}^\infty \alpha_{-n}^i  \alpha_n^i - a\hbar $ and $\tilde N_{(0)} = \sum_{n=1}^\infty \tilde\alpha_{-n}^i \tilde\alpha_n^i - a\hbar$. Standard arguments tell us that $a = \frac{D - 2}{24}$. In \cite{PNString} we work out the Poisson algebra of the Noether charges of the global symmetries of the NLO Polyakov action with a flat target space. We expect the quantum theory to have the same symmetry algebra if we choose $D=26$ in line with the results of \cite{Gomis:2000bd}.\\

\noindent\textbf{Discussion}.

The beta functions for NR string theory have been computed in \cite{Gomis:2019zyu,Gallegos:2019icg,Yan:2019xsf} and an action has been proposed that reproduces all but one (the string analogue of the Poisson equation) of the beta functions in \cite{Bergshoeff:2021bmc}. Based on the action for NR gravity obtained using the particle $1/c^2$ expansion of GR \cite{Hansen:2018ofj,Hansen:2020pqs}, we expect that the string $1/c^2$ expansion of NS-NS gravity could lead to an action principle for all the beta functions of NR string theory.

Recent studies of non-relativistic string theories, including other works such as \cite{Harmark:2018cdl,Bagchi:2021rfw}, point to the existence of a landscape of string theories beyond the Lorentzian ones we are familar with. It would be of interest to study open string sectors and D-brane like objects in such theories (see e.g.~\cite{Gomis:2020fui,Blair:2021ycc}). In this light, it would be interesting to apply the $1/c^2$ expansion to the study of open strings and D-brane actions.

\noindent\textbf{Acknowledgments.} We thank Leo Bidussi, José Figueroa-O'Farrill, Troels Harmark, Niels Obers, and Gerben Oling for useful discussions. The work of JH is supported by the Royal Society University Research Fellowship ``Non-Lorentzian Geometry in Holography'' (grant number UF160197). The work of EH is supported by the Royal Society Research Grant for Research Fellows 2017 “A Universal Theory for Fluid Dynamics” (grant number RGF$\backslash$R1$\backslash$180017).
\providecommand{\href}[2]{#2}\begingroup\raggedright\endgroup

\end{document}